\documentclass[a4paper]{article}
\usepackage{color}
\usepackage{setspace}
\usepackage{float}
\usepackage{epstopdf}
\usepackage{standalone}
\usepackage{amsmath}
\usepackage{pdflscape}
\usepackage{authblk}
\usepackage{graphicx}
\usepackage[flushleft]{threeparttable}
\usepackage{natbib}
\usepackage[citecolor=blue]{hyperref}

\usepackage[margin=1.1in]{geometry}

\newcommand{\interior}[1]{%
 {\kern0pt#1}^{\mathrm{o}}%
}
\usepackage [english]{babel}
\usepackage [autostyle, english = american]{csquotes}
\MakeOuterQuote{"}

\usepackage{algorithm,algpseudocode}
\usepackage{caption}
\usepackage[makeroom]{cancel}

\title{\LARGE{\textbf{An Ensemble Approach to Predicting the Impact of Vaccination on Rotavirus Disease in Niger}}}
\author[1]{Jaewoo Park}
\author[2]{Joshua Goldstein}
\author[1]{Murali Haran\thanks{Email addresses: joshg22@vbi.vt.edu (J. Goldstein), jzp191@psu.edu (J. Park), muh10@psu.edu (M. Haran), mjf283@psu.edu (M. Ferrari)}}
\author[3]{Matthew Ferrari}
\affil[1]{\footnotesize Department of Statistics, The Pennsylvania State University, University Park, PA
16802, USA}
\affil[2]{\footnotesize Social and Data Analytics Laboratory, 900 N Glebe Rd, Virginia Tech, Arlington, VA 22203. USA}
\affil[3]{\footnotesize Department of Biology, The Pennsylvania State University, University Park, PA
16802, USA}
\begin{document}

\maketitle
\begin{abstract}
Recently developed vaccines provide a new way of controlling rotavirus in sub-Saharan Africa. Models for the transmission dynamics of rotavirus are critical both for estimating current burden from imperfect surveillance and for assessing potential effects of vaccine intervention strategies. We examine rotavirus infection in the Maradi area in southern Niger using hospital surveillance data provided by Epicentre collected over two years. Additionally, a cluster survey of households in the region allows us to estimate the proportion of children with diarrhea who consulted at a health structure. Model fit and future projections are necessarily particular to a given model; thus, where there are competing models for the underlying epidemiology an ensemble approach can account for that uncertainty. We compare our results across several variants of Susceptible-Infectious-Recovered (SIR) compartmental models to quantify the impact of modeling assumptions on our estimates. Model-specific parameters are estimated by Bayesian inference using Markov chain Monte Carlo. We then use Bayesian model averaging to generate ensemble estimates of the current dynamics, including estimates of $R_0$, the burden of infection in the region, as well as the impact of vaccination on both the short-term dynamics and the long-term reduction of rotavirus incidence under varying levels of coverage. The ensemble of models predicts that the current burden of severe rotavirus disease is 2.9 to 4.1\% of the population each year and that a 2-dose vaccine schedule achieving 70\% coverage could reduce burden by 37-43\%. 

\end{abstract}

\noindent%
{\it Keywords: Rotavirus, Vaccine, Niger, Susceptible-Infectious-Recovered models, Bayesian model averaging} 
\vfill

\section{Introduction}

~~~~Diarrheal disease is the second leading cause of death around the world for children under 5 years of age \citep{black2010global}. Though there are many infectious causes of diarrheal disease in children, rotavirus is the leading cause of gastroenteritis \citep{bryce2005estimates,unicef2010diarrhoea,tate2012}.
In many countries, better sanitation, hygiene and access to care have reduced the burden of diarrhea \citep{clasen2007interventions, kilgore1995trends}. 
Despite this trend, the proportion of diarrheal hospitalizations attributable to rotavirus  increased between 2000 and 2004 \citep{parashar2006rotavirus}. The recent development of new prophylactic vaccines for rotavirus is a promising advance in the prevention of diarrheal disease and the reduction of overall childhood mortality \citep{patel2009association,madhi2010effect}.

Observation of rotavirus dynamics and estimation of the burden of rotavirus disease is limited both by non-specific surveillance and under-reporting. The dynamics of rotavirus transmission must often be inferred from non-specific temporal surveillance of diarrheal disease that includes multiple causes. This is analogous to the dynamics of specific influenza strains, which are commonly inferred from non-specific time series surveillance of influenza-like illness (ILI) that includes infection by multiple influenza strains (influenza A and B), as well as additional viral infections, for example parainfluenza, coronavirus, rhinovirus \citep{riley2003transmission, chowell2011characterizing}. In sub-Saharan Africa, the cause of diarrheal disease is often unknown due to a lack of diagnostic capacity \citep{mwenda2010burden}. Even when the cause of disease is known, an unknown fraction of cases will occur in the community and never be recorded by the health system, leading to a potentially significant level of under-reporting. Dynamic models in general, and so-called state-space models in particular, have been an important tool in the assessment of disease burden from non-specific or imperfect surveillance \citep{ionides2006inference,ferrari2008dynamics,breto2009time}. 
We estimate the burden of rotavirus in the Maradi region of Niger by synthesizing two sources of data. We use hospital surveillance data collected by Epicentre for the incident cases over time, including lab-confirmation to assess the likelilhood that a case of severe diarrhea is caused by rotavirus. In addition, we use a cluster survey of households conducted to estimate the proportion of diarrheal disease cases in the region seeking care. The latter is used to help  estimate the reporting rate. 

State-space models rely on the temporal correlation in a dynamic model to make the unobservable true state of the system, that is, the incidence of the pathogen of interest, estimable from noisy or imperfectly sampled data \citep{jones1993longitudinal}. Thus, the inference about disease burden is conditional on the structure of the underlying dynamic model. For pathogens with well characterized epidemiology, such as measles and influenza, the application of state-space models to infer disease burden and transmission dynamics has become common \citep{ionides2006inference, cauchemez2008likelihood, breto2009time, simons2012assessment}. The dynamics of rotavirus, which itself comprises multiple strains that result in varying levels of cross-protective immunity to other strains, has been variously described by a suite of different models \citep{pitzer2012direct}.
Therefore, inference about rotavirus burden is limited both by imperfect surveillance of rotavirus infection and uncertainty about the underlying transmission dynamics. Rather than condition our analysis on any one model, we fit the observed time series to a suite of 5 different model structures and assumptions to account for uncertainty in model parameters as well as the dynamics represented in the models themselves.

While the development of several novel rotavirus vaccines is a promising advance for the control of diarrheal disease in children, the potential impact of the introduction of these vaccines at the population-scale is uncertain. The predicted impact of vaccine introduction may depend both on the efficacy of the vaccine and model structure; for example \citep{pitzer2012direct} proposed alternative models for boosting of immunity following sequential exposure to rotavirus. Bayesian model averaging (BMA) \citep{bates1969combination,hoeting1999bayesian} allows for the integration of predictions of multiple models, weighted by their posterior support, to generate a single ensemble estimate that accounts for uncertainty in model selection. Here, via BMA, we use the ensemble of fitted models to predict the short-term and long-term impact of vaccination on rotavirus incidence. We then estimate the predicted impact using the vaccine efficacy from two different studies. Our ensemble approach predicts that the current burden of severe rotavirus disease is 2.9 to 4.1\% of the population each year and that a 2-dose vaccine schedule achieving 70\% coverage could reduce burden by 37-43\%. 

\section{Material and Methods}

~~~~We use data from two sources: a time series of clinic admissions for diarrheal disease and  a community based survey of health-seeking behavior. Clinic surveillance covers a collection of health centers and district hospitals from four districts in the Maradi region of Niger including Aguie, Guidan Roumdji, Madarounfa, and the city of Maradi. A total of 9,590 cases of diarrhea in children under 5 were recorded from December 23, 2009 to March 31, 2012 (118 weeks). For each patient age in months, village of origin, date of consultation were recorded. 
Also noted were potential symptoms including temperature, duration of diarrhea before consultation, presence of blood in the stool, presence and duration of vomiting, and level of dehydration. In each case a rapid test was administered for detecting rotavirus. 2,921 cases tested positive for rotavirus via the rapid test. A subset of 378 cases testing positive for rotavirus were also genotyped. While 32 separate strains were identified, more than two thirds of positive cases were of strains G12P[8] or G2P[4].

The distributed nature of Niger's healthcare system is a challenge for surveillance. Roughly a third of all health centers in these districts were included. Notably absent were the many local health posts staffed by community health workers. To estimate both the fraction of cases seeking care at a health center, and the fraction seeking any level of care, a second source of data is needed. We use a community survey \citep{page2011health} in the region of children under 5 to get estimates of these reporting rates.

A total of 2940 children under 5 were selected for inclusion in the cluster survey from households across the four districts. Clusters were allotted according to the population of each village from census data. Sampling weights accounted for household composition and the relative populations of the districts. Among those surveyed, 1099 caregivers reported at least one episode of diarrhea during the recall period of 27 days. Respondents reported whether they sought care at a health structure. We use the reporting rate of severe diarrhea, which is defined as the presence of acute watery diarrhea and the presence of two or more of the signs of loss of consciousness, sunken eyes, and an incapacity to drink or drinking very little.

From the cluster survey we determine that an estimated total of 42.9\% of caregivers who reported severe diarrhea consulted at a health center $\left(95\% \text{CI}: (33.1\%,52.7\%)\right)$. The rest either sought care at a district hospital, local health post or did not seek care at a formal health structure. This estimate is used as a proxy for the reporting rate of rotavirus. More specifically, this information is used to construct an informative prior for our Bayesian approach (as described in the supplementary material).

\subsection{Model Overview}

~~~~We consider a range of dynamic models for rotavirus transmission. Information linking individual-level data on the course of infection to the between-person transmission of rotavirus is lacking, leading to variation in the structure of mathematical models for rotavirus \citep{pitzer2012direct}. Using a range of different models allows us to account for the uncertainty in estimation due to model choice. The five models we consider are SIR-like compartmental models of transmission, building upon the models in \cite{pitzer2012direct}. 
We incorporate age into the model with separate compartments for ages from 0-1 month, 2-3 months, 4-5 months, 6-11 months, 12-23 months, and 24-59 months. 
Fixed parameters are estimated from England and Wales data as described in \cite{pitzer2012direct}. 

Here we very briefly outline the main features of five models, Models A through E, based on the SIR framework. Details of the model and inferential procedure are described in the supplementary material. Model A tracks severe and mild rotavirus separately. Severe infections have larger force of infections than mild infections. Unlike Model A, Models B-E assume successive infections and immunity are obtained through
repeated infections. Subsequent infections will have a reduced susceptibility to infection and level of infectiousness. Model C allows for an incubation period of infections as well. In Model D there is no temporary immunity during successive infections and immunity is granted after all repeated infections. Model E assumes that full immunity can be obtained during successive infections. All model parameters are estimated via Markov chain Monte Carlo (MCMC) and estimated burden over time were obtained from each model.

\subsection{Vaccination}

~~~~We assume vaccination imparts immunity comparable to a natural infection, and consider a strategy wherein a first dose is administered at 2 months of age and a second dose is administered at 4 months. The vaccine was assumed to confer protection comparable to protection conferred by primary infection following the first dose. The second dose confers additional protection comparable to that conferred by secondary infection. For Model A, where the risk of infection does not decrease based previous number of infections, a separate input parameter is used for the vaccine efficacy. The vaccine efficacy is set to be equal to the predicted efficacy for Models B-E (see supplementary material for details). We study the effect of the vaccine under varying levels of coverage. The short-term effect of vaccination is assessed by looking at incidence over a five year period following introduction of the vaccine. The long-term effect is measured by the yearly reduction in incident cases of Rotavirus gastroenteritis (RVGE) measured 20 years after introduction of the vaccine. Field efficacy of a multi-dose rotavirus vaccination strategy is uncertain.  To reflect this uncertainty, we investigate the impact of vaccination using the value of efficacy from two  different studies. First based on the results of \citep{lopman2012understanding}, for low income countries, we assume a seroconversion rate of 63\%. Second, a recent study of a 3-dose vaccination strategy in Niger \citep{isanaka2017efficacy} estimated efficacy of 66.7\% with all doses. The details of representing these two estimates of efficacy in the 5 models are presented in the supplement.

\section{Results}

~~~~We fit each model independently and estimate parameters. Then we calculate ensemble estimates using Bayesian model averaging (BMA) \citep{bates1969combination,hoeting1999bayesian} to formalize uncertainty in model selection (see supplementary material for details). Posterior model probabilities (PMP) measures how much each model is supported by the data. Following the BMA approach, based on these probabilities, we provide a weighted average of estimates from five different models. There is significant discordance across models in the measures of model fit (Table \ref{tab:BMAresults}). Model C, the model with incubation periods performs the best. Notably, Model A, the only model that does not allow for successive infections with decreased levels of infectiousness, performs significantly worse as measured by posterior model probability. 

\begin{table}[hh]
\centering
\begin{tabular}{cccc}
\hline
Model & Probability & $R_{0}$ & Burden\\
\hline
A  &  0     &  30.7 (25.8,34.3)  &  9.2 (8.1,10.1)  \\
B  &  0.01  &  13.9 (12.7,15.4)  &  3.5 (3.1,3.8)   \\
C  &  0.92  &  13.4 (11.7,15.3)  &  3.5 (3.1,3.9)   \\
D  &  0.03  &  11.2 (9.4,12.7)   &  3.6 (3.2,4.1)   \\
E  &  0.04  &  10.3 (9.5,12.6)   &  3.2 (2.9,3.5)   \\
BMA  &      &  13.4 (10.3,15.4)  &  3.5 (2.9,4.2)   \\
\hline
\end{tabular}
\caption{For each model we provide posterior model probability (PMP), the basic reproductive number $R_{0}$, and estimated burden. Burden corresponds to yearly cases with severe RVGE (\% of population). The last row corresponds to the model-averaged (via Bayesian model averaging) versions of these estimates.}
\label{tab:BMAresults}
\end{table}

\subsection{Pre-vaccination}

\begin{figure}
\includegraphics[width=\linewidth]{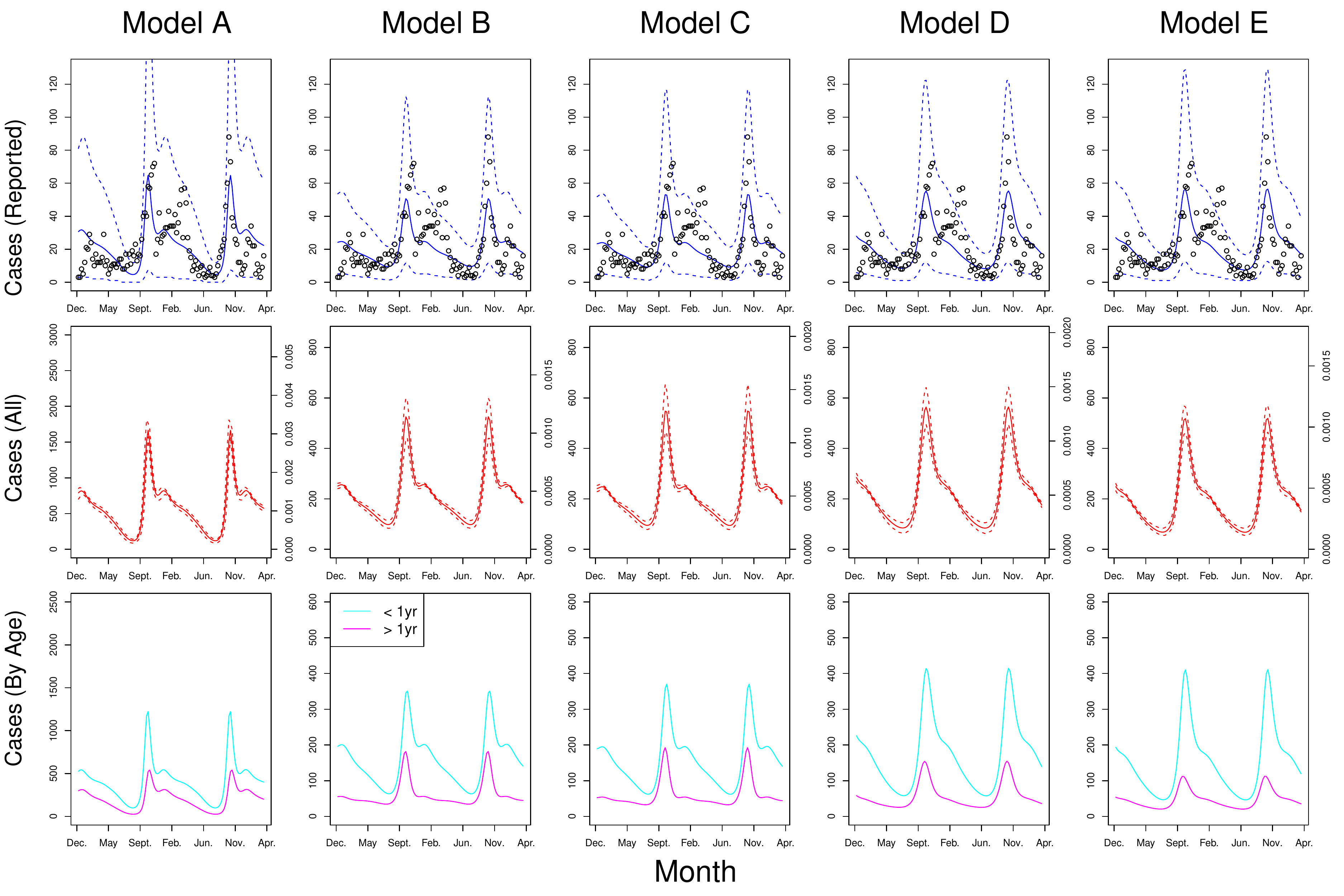}
\caption{Burden estimates under the five fitted models. Dashed lines denote 95\% confidence interval. Top: weekly reported cases of RVGE and model projections. Middle: model projections of all severe RVGE cases. Bottom: model projections of RVGE incidence by age. Lines are model projections while points represent observed cases.}
\label{fig:burden}
\end{figure}

~~~~Our fitted models allow us to construct estimates of the burden in these four districts (Table \ref{tab:BMAresults}). Of children under five, an approximate 3.5\% per year develop severe RVGE as estimated by Models B-D, though this estimate is lower for Model E and significantly larger for Model A. The basic reproductive number $R_0$ is found as the largest eigenvalue of the next-generation matrix \citep{diekmann1990definition} and significantly larger for Model A. BMA for burden and $R_{0}$ are close to those of Model C, which has the highest weight. In Figure \ref{fig:burden}, we plot our model projections with uncertainty for reported cases of rotavirus as well as for all cases of severe RVGE.  
We also note that Models B-E predict a steep decline in cases in children over 1y of age following the epidemic peak; cases in infants under 1y, by contrast, are predicted to decline more slowly. 

\begin{figure}
\includegraphics[width=\linewidth]{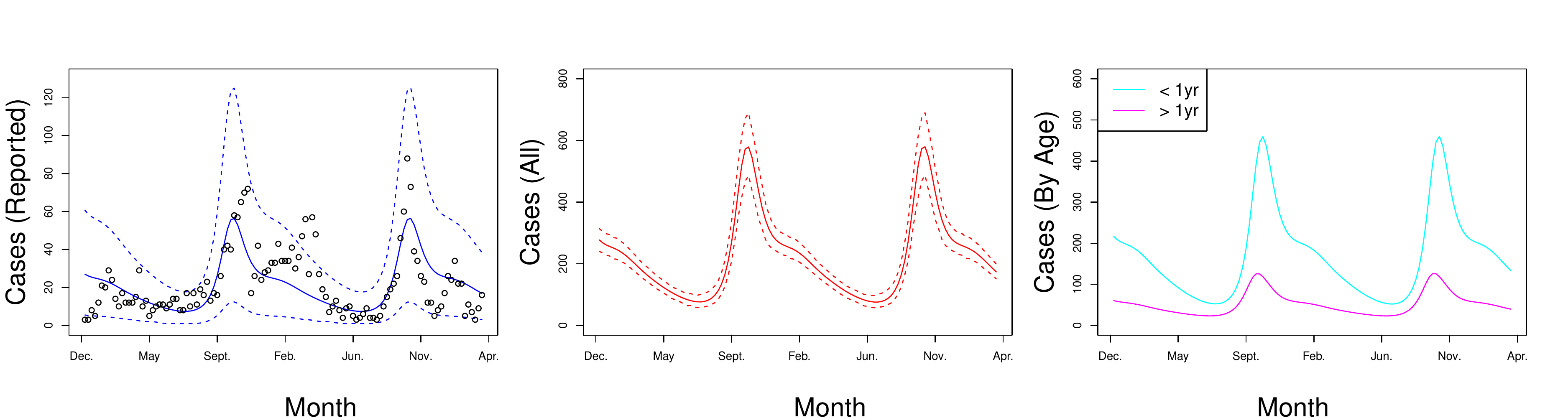}
\caption{Model-averaged (BMA) burden estimates from the five fitted models. Dashed lines denote 95\% confidence interval. Left: weekly reported cases of RVGE and model projections. Middle: model projections of all severe RVGE cases. Right: model projections of RVGE incidence by age. Lines are model projections while points represent observed cases.}
\label{fig:BMAburden}
\end{figure}

Figure \ref{fig:BMAburden} shows the BMA-based model projections which are close to those of Model C. However we note that BMA-based projections have wider confidence intervals because averaged projections incorporate model uncertainty.

All of the fitted models are able to successfully capture the observed age distribution of cases (Figure \ref{fig:BMAageDisttry2}), though Models C and E predict noticeably more cases than observed for older children (2-5 years). The models vary in their ability to capture the temporal dynamics. During the second year of hospital surveillance we can see a secondary peak in the number of cases that is not captured by our fitted model, although we did find that the model dynamics can produce this double peak through an interaction of a high birth rate and seasonal variation when the seasonal forcing is stronger than that estimated here. BMA shows a similar trend as the Model C, which has the highest weight.

\subsection{Projected Impact of Vaccination}

\begin{figure}
\centering
\includegraphics[width=.7\linewidth]{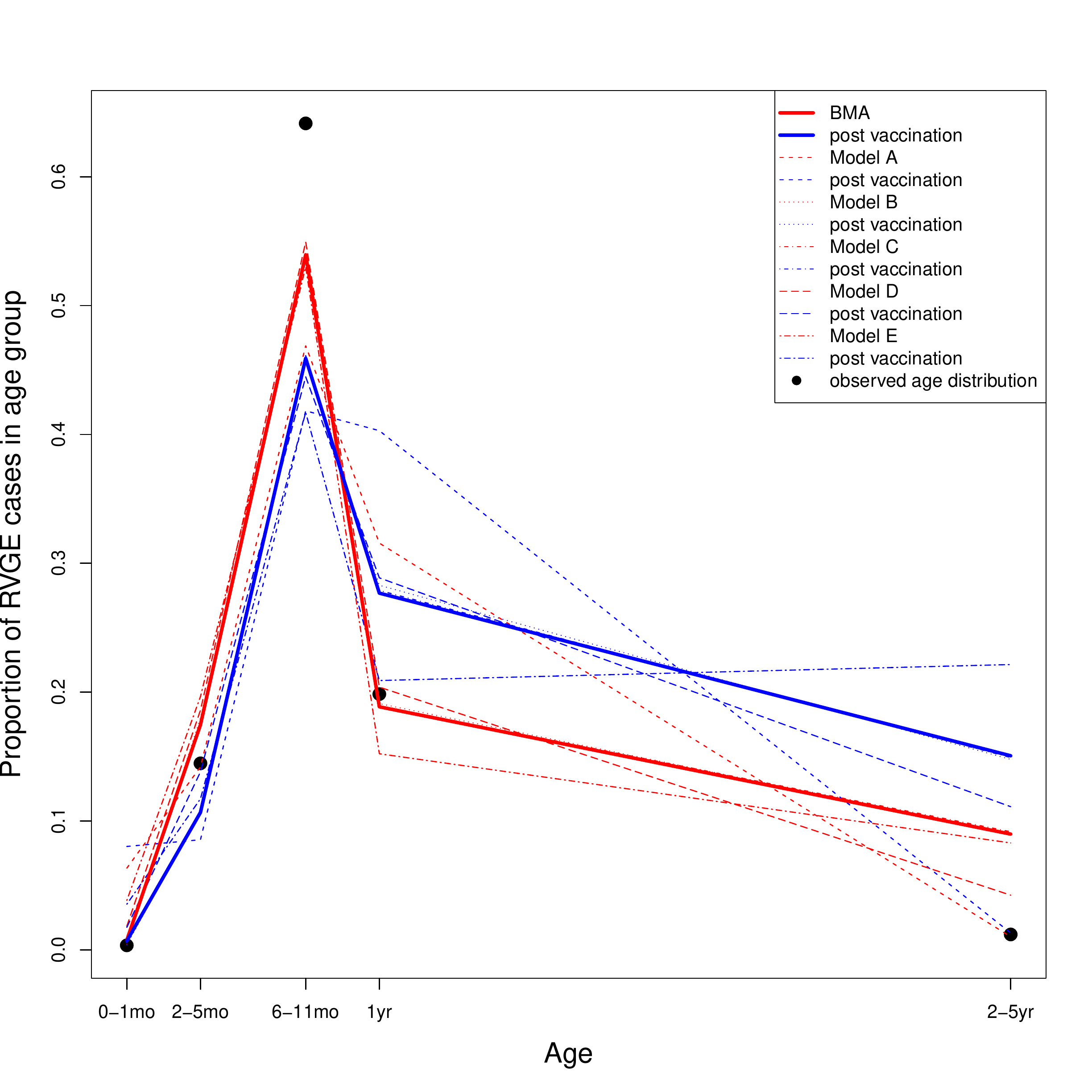}
\caption{Distribution of cases across age groups observed in the data (black dots), predicted by the models (red lines), and predicted after vaccination has been introduced at 70\% coverage (blue lines).}
\label{fig:BMAageDisttry2}
\end{figure}

~~~~Here we investigate the impact of vaccination based on the seroconversion rate for low socio-economic settings \citep{lopman2012understanding}. In the supplementary material we provide the impact of vaccination using a  different value of efficacy which is measured based on a 3-dose strategy \citep{isanaka2017efficacy}. This was qualitatively similar, but quantitatively small  compared to the results in the main paper. Vaccination causes a noticeable shift in the age distribution across Models B-E (Figure \ref{fig:BMAageDisttry2}), with a higher proportion of RVGE cases occurring in older children. This has significant benefits when considering the age-specific mortality of rotavirus is higher for children under 2 years of age \citep{morris2012rotavirus}. The BMA-based burden shows a similar trend. 

\begin{figure}
\includegraphics[width=\linewidth]{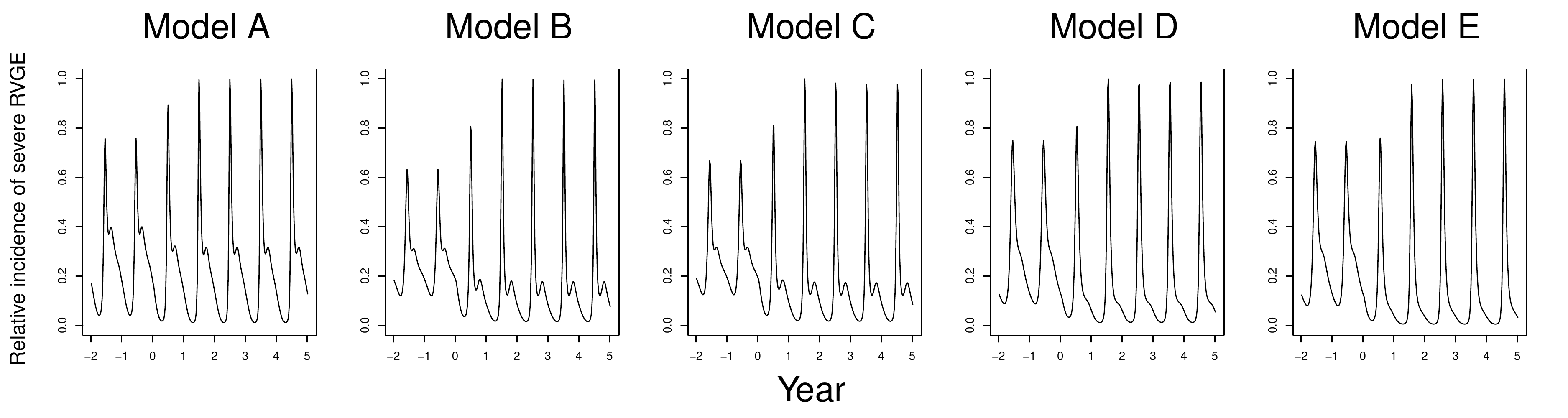}
\caption{Relative incidence of severe RVGE after vaccination has been introduced into the models assuming 70\% coverge, out to five years after vaccination has been introduced. The vaccination has been introduced at 0 year. }
\label{fig:vaccShortTerm}
\end{figure}

Over the short term, Models A-E predict an overall decline in total burden, but an increase in the magnitude of peak incidence (Figure \ref{fig:vaccShortTerm}). 

\begin{figure}
\includegraphics[width=\linewidth]{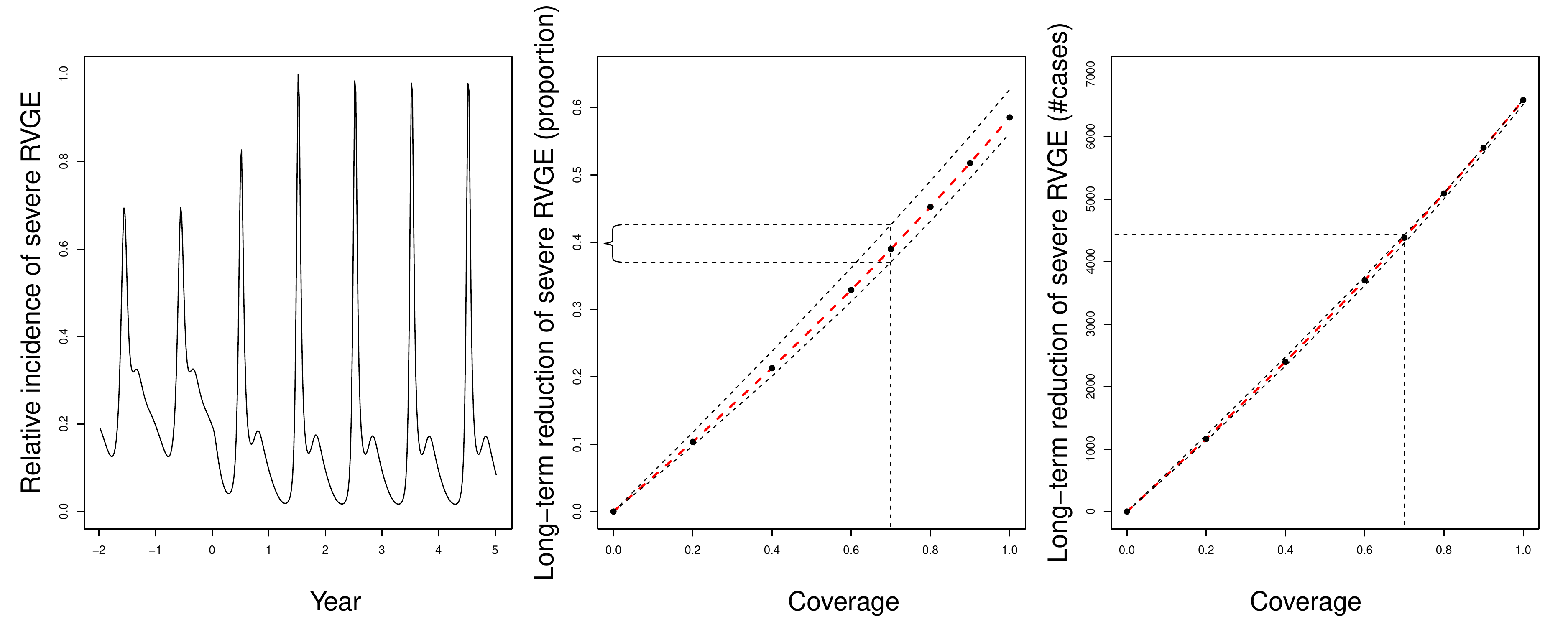}
\caption{Relative incidence of severe RVGE (Left), percent (Middle) and absolute (Right) long term reduction in cases by coverage for Bayesian model averaging from the five fitted models. Dashed lines denote 99\% confidence interval. The vaccination has been introduced at 0 year. Variation in reduction for a fixed (70\%) level of coverage is demonstrated.}
\label{fig:BMAvaccReduce}
\end{figure}

Figure \ref{fig:BMAvaccReduce} provides short term and long term impact of vaccination which are model-averaged values from five different models. The short term trend of vaccination impacts based on BMA is similar to that of Model C.  At equilibrium (long term), we can observe the reduction in severe rotavirus cases with higher levels of coverage. For a fixed (70\%) level of coverage, we predict 38.9\% reduction of severe RVGE $\left(99\% \text{CI}: (37.1\%,42.6\%)\right)$ over the long-term. Based on the recent vaccine efficacy study in \cite{isanaka2017efficacy}, we predict 29.6\% reduction $\left(99\% \text{CI}: (28.0\%,32.7\%)\right)$ in RVGE over the long-term. Details are provided in the  supplementary material.

\section{Discussion}

~~~~Diarrheal disease is a major source of childhood morbidity and mortality. However, the multi-etiology nature of diarrheal disease means that it is difficult, in the absence of lab confirmation, to infer total burden or project the consequences of novel interventions. We have rich but short-term data with which to understand the dynamic process; 
in combination with survey data on health-seeking behavior; however, we can bring additional information to bear on the observation rate to interpret the patterns from the non-specific clinic surveillance.

For rotavirus, the uncertainty inherent in imperfectly observed incidence is compounded by the lack of a generally accepted model and debate about the underlying mechanisms that drive the epidemiological response \citep{pitzer2012direct}. This motivates an ensemble approach, using a combination of different models along with quantitative surveillance to get practical measures of burden and projections about the operational impact of controls.  
This multi-model ensemble approach is common in geosciences \citep{mcavaney2001model, stainforth2005uncertainty, tebaldi2005quantifying}, 
where different assumptions on complex underlying processes can produce different climate projections, which motivates a probabilistic forecast from a variety of models. A competing models approach has been adapted to epidemiological problems as well, such as choosing an optimal strategy for measles vaccination \citep{shea2014adaptive} and assessing the impact of control actions for foot and mouth disease outbreaks \citep{lindstrom2015bayesian}.

Here we formally address these two sources of uncertainty, using a state-space model to address the problem of incidence from non-specific surveillance data, and comparing the inference from an ensemble of proposed models to address the uncertainty in the underlying dynamics. Our ensemble approach suggests robust support for some general patterns of rotavirus dynamics. The peak transmission is well estimated, with a maximum in early March, with little variation between models. Rainfall, which is a primary driver of seasonality in the region, peaks in August.  
\cite{bharti2011explaining} found that early March, when urban population density is at its maximum due to seasonal rural-urban migration, was the peak season for transmission of measles. Though measles is transmitted through aerosolized droplets, the similarity in the peak seasonality suggests that higher population density may also facilitate transmission of rotavirus.

We find the SEIRS structure in Model C (model with incubation period) best explains the observed data. In this model, subsequent infections have decreased levels of infectiousness and lower risk of infection compared to the initial infection. All models except for Model A, which offers the worst fit to the data, include this dynamic. The estimated basic reproductive number is fairly robust across Models B-E. In particular, point estimates for models B-E vary from 10.3 to 13.9 in Table \ref{tab:BMAresults}, though Model A has a much larger $R_0$.

There is an observed double peak in incidence (Figure \ref{fig:burden}) during the second year of observation which our fitted models do not capture. However, this may be an anomaly, as the double peak is not seen strongly during the first and third years. We note that our models are capable of reproducing this behavior when the seasonal variation in transmission is stronger than the best fit estimate, via an interaction between seasonal effects and the high birth rate in the region. More complex explanations for such double peaks have been observed elsewhere. In cholera, similar to rotavirus in transmission, local ecological variations were responsible for bimodal incidence \citep{de2011role}.

Our estimate of overall burden of severe RVGE is robust across Models B-E. In spite of the fact that the full epidemiological processes are unknown, we can be fairly sure that the total yearly burden among children under 5 is in the vicinity of 3.5\% (Table \ref{tab:BMAresults}). Model A predicts a 3-fold greater incidence of severe RGVE; however, this model has the weakest support and model-averaged burden is similar to Models B-E.

While uncertainty in retrospective dynamics and disease burden can be characterized using different models, additional uncertainty about the efficacy of proposed interventions limits the ability to predict future dynamics and disease burden.  \cite{atherly2012projected} estimated that rotavirus vaccine could result in 2.46 million childhood deaths between 2011-2030.  Of course, uncertainty in the seroconversion rate \citep{lopman2012understanding} and achievable vaccination coverage means that the true benefit of these vaccines is unknown. Here, we used the ensemble prediction to project the potential impact of rotavirus vaccination in the Niger setting under two scenarios for vaccine efficacy; thus integrating both dynamic uncertainty due to different models and sensitivity to the realized effectiveness of a vaccine program.  Using a vaccine efficacy derived from \cite{lopman2012understanding} we estimate that 70\% coverage could result in 37-43\% reduction in severe RVGE in children under 5.  \cite{isanaka2017efficacy} reported a lower efficacy from a 3-dose schedule in Niger; this would lower the projected reduction of severe RVGE to 28-33\%.  Notably, although BMA estimates a total reduction in yearly cases using both the efficacy reported in \cite{lopman2012understanding} and \cite{isanaka2017efficacy}, it also predicts higher peaks where more cases are observed than pre-vaccination. This short-term difference in cycle amplitude for these models is a phenomenon anticipated by \cite{pitzer2009demographic}. Anticipation of this shift in dynamic regime caused by vaccination may be critical to the interpretation of short-term surveillance as the observation of higher peak incidence following the introduction of vaccination may be wrongly interpreted as a failure in the vaccination program.

Dynamic models are a powerful tool to interpret disease surveillance data and anticipate the potential consequences of interventions. The method we describe here addresses two main sources of uncertainty: imperfectly observed data and scientific uncertainty about epidemiological dynamics.  Our methods also allow us to  identify key epidemiological interpretations -- transmission seasonality and the proportional impact of vaccination -- that are robust to model choice, and those that are model dependent, that is, $R_0$ and the annual burden of severe RVGE.  By assessing the fit of the observed surveillance to each model, we find that these latter measures are robust within the subset of well supported models.   

\section{Acknowledgments}

~~~~The authors are grateful to  Epicentre for providing the data sets for this research project. MF is funded by a grant from the Ecology and Evolution of Infectious Disease program of the NSF/NIH (award number 1 R01 GM105247-01).

{\it Potential conflict of interest statement:} none of the authors has conflicts of interest.
\clearpage

\appendix
\begin{center}
\title{\LARGE\bf Supplementary Material for An Ensemble Approach to Predicting the Impact of Vaccination on Rotavirus Disease in Niger}\\~\\
\author{Joshua Goldstein, Jaewoo Park, Murali Haran, and Matthew Ferrari}
\end{center}

We provide details below about the five different dynamic models along with information about computational methods used to perform inference for each of them. In addition, we describe the implemention of the Bayesian model averaging approach used in the manuscript. 

\appendix
\section{Model Details}

\begin{figure}[h!]
\centering
\includegraphics[width=5in]{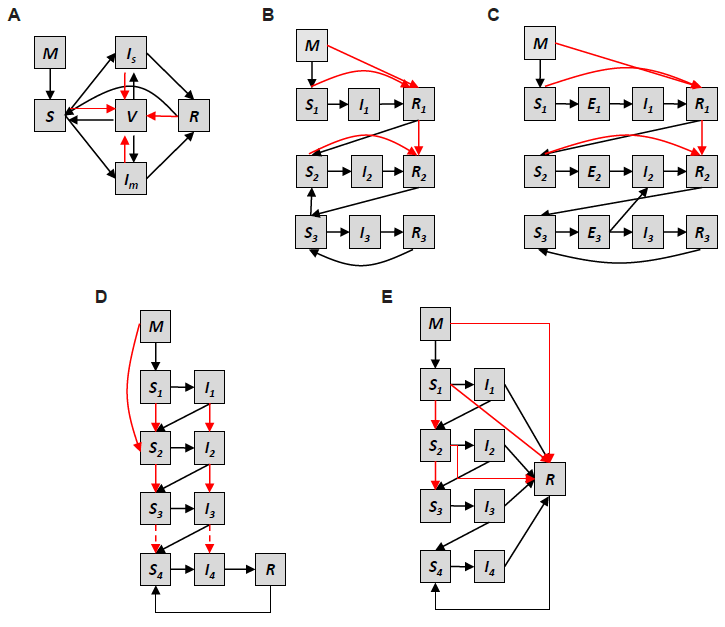}
\caption{Structure of the compartmental models adapted from \cite{pitzer2012direct}.}
\label{modelstructure}
\end{figure}

~~~~The structure of the models is given in Figure \ref{modelstructure}, which we explain in detail below. Common to each of the models we describe, we assume a time-varying transmission rate with a period of one year to account for seasonality, 
\[
\beta_i(t) = \beta_{0i} \left( 1 + b \hspace{1mm} \text{cos} \left( \dfrac{ 2 \pi t - 52 \phi } { 52 } \right) \right), 
\]
where $t$ is time in weeks, $\beta_{0i}$ is the baseline rate for age class $i$, and $b$ and $\phi$ are the amplitude and offset of the seasonal variation.

We also assume the birth rate $\mu(t)$ varies with time. The mean weekly birthrate is estimated by $\bar{\mu} =  1 / (5 \times 52)$. The variation in monthly birth rate is shown in Table \ref{tab:birthrate}. Finally, for each model we assume a negative binomial observation process with mean equal to the number of weekly reported cases and dispersion parameter $r$.

\begin{table}[h]
\centering
\caption{Seasonal variation in birth rate in Niger, estimated from 1980-2000 using Demographic and Health Surveys. \cite{dorelien2013time} An amplitude of $-.17$ for January tells us the birth rate is $17\%$ below the mean.}
\begin{tabular}{c|cccccccccccc}
\hline
Month & Jan & Feb & Mar & Apr & May & Jun & Jul & Aug & Sep & Oct & Nov & Dec \\ 
\hline
Amplitude & -.17 & .01 & .03 & .25 & .12 & .03 & -.01 & .09 & .01 & .13 & -.31 & -.17 \\ 
\hline
\end{tabular}
\label{tab:birthrate}
\end{table}

We describe in detail the dynamics of each of the five models outlined in Figure \ref{modelstructure}. Model A \citep{shim2006age,atkins2012impact} is an SIRS model in which severe and mild rotavirus are tracked separately. Severe infections have a longer duration and contribute more to the overall force of infection. Following infection, there is a period of temporary immunity that wanes over time. The model is age structured with age groups 0-1 month, 2-3 months, 4-5 months, 6-11 months, 1 year, and 2-5 years indexed by $i$. The differential equations describing the model dynamics are:

\begin{align}
& \phantom{\dfrac{dM_i}{dt} =  \alpha_{i-1} M_{(i-1)}} \text{Model A} \notag \\
& \dfrac{dM_i}{dt} = \alpha_{i-1} M_{(i-1)} - \alpha_i M_i + \mu N - \delta M_i \\
& \dfrac{dS_i}{dt} = \alpha_{i-1} S_{(i-1)} - \alpha_i S_i + \delta M_i - \lambda_{i} S_i + \tau R_i \\
& \dfrac{dI_s,i}{dt} = \alpha_{i-1} I_{s,(i-1)} - \alpha_i I_{s,i} + \lambda_{s,i} S_i -  \gamma_s I_{s,i} \\
& \dfrac{dI_m,i}{dt} = \alpha_{i-1} I_{m,(i-1)} - \alpha_i I_{m,i} + \lambda_{m,i} S_i -  \gamma_m I_{m,i}
\end{align}

 Movement between age classes occurs at rates dependent on the length of the interval in weeks, $\boldsymbol\alpha = \left\{ \dfrac{1}{8}, \dfrac{1}{8}, \dfrac{1}{8}, \dfrac{1}{24}, \dfrac{1}{48}, \dfrac{1}{144} \right\}$. The force of infection for age class $i$ is given by
$\lambda_i =  \displaystyle\sum_{j=1}^6 \beta_j(t) C_{ij} \dfrac{ (I_{s} + 0.5 I_{m}) }{ N_j }$,
assuming that relative infectiousness for mild infections is less than for severe RVGE. Here $C_{ij}$ represents the frequency of contact from age class $i$ onto class $j$ \citep{hethcote1996modeling}, and satisfies $f_i C_{ij} = f_j C_{ji}$ where $f_i$ is the fraction of the population in class $i$. We make the simplifying assumption that contact between age groups is homogeneous.
With the absence of data on rotavirus infections for children over 5 and adults, we also assume the population of children under 5 is closed and consider child-child transmission only. Infection with rotavirus is typically asymptomatic \citep{pitzer2012direct} or unreported for older children and adults, but could potentially play a role in transmission.
The contact matrix is

\[ \small C = \left( \begin{array}{cccccc}
1 & 1 & 1 & 1 & 1 & 1 \\
1 & 1 & 1 & 1 & 1 & 1 \\
1 & 1 & 1 & 1 & 1 & 1 \\
3 & 3 & 3 & 1 & 1 & 1 \\
6 & 6 & 6 & 2 & 1 & 1 \\
18 & 18 & 18 & 6 & 3 & 1
\end{array} \right) \]. 

The differences in our age groups means that the contact matrix is not symmetric, for example we assume the population from 2-5 years is 18 times larger than the population from 0-1 months. 

After a period of maternal immunity $(M_i)$, individuals can be susceptible $(S_i)$, infected with either mild $(I_{m,i})$ or severe $(I_{s,i})$ rotavirus, or recovered $(R_i)$. 
These represent the number of individuals in each class. In (1) we see how the number of children protected by maternal immunity change over time. Newborns are added to this class at rate $\mu$ and individuals leave this class when maternal immunity wanes with rate $\delta$, where the mean period of maternal immunity is assumed to be 13 weeks ($\delta = \frac{1}{13}$).

When maternal immunity wanes children are susceptible to rotavirus infection. 
In (2), we see that individuals enter the susceptible class when maternal immunity wanes. 
They become infected at a rate given by the force of infection $\lambda_i$. 
After recovery, individuals may reenter the susceptible class at rate $\tau$, where the mean period of immunity following infection is fixed at one year ($\tau = {1}/{52}$).

Equation (3) models the change in total infections with severe rotavirus. 
We assume the proportion of infections with severe rotavirus is lower than mild by setting $\lambda_{s,i} = 0.24 \lambda_i$. 
Individuals leave the infected with severe rotavirus for a mean period of one week ($\gamma_s = 1$) following which they are considered to be recovered. 
Similarly, (4) tracks the total infections with mild rotavirus, with $\lambda_{m,i} = 0.76 \lambda_i$ and a mean infectious period of just half a week ($\gamma_m = 2$).

Only a fraction of infections with rotavirus develop RVGE (fixed at 24\%), and we assume only severe cases are reported, so the expected number of reported cases for age class $i$ is given by $0.24 \rho \lambda_i S_i$ where $\rho$ is the reporting rate. We make the simplifying assumption for all models that $\rho$ is constant across time and does not vary by age group.

Model B \citep{pitzer2009demographic} is an SIRS model allowing for successive infections in which a second, third or subsequent infection will have a reduced susceptibility to infection and level of infectiousness. 
This represents partial immunity granted through repeated infections. 
Only a fraction of individuals in the a first or second infectious class are assumed to develop severe RVGE. 
The model dynamics are described by as follows.

\begin{align*}
& \phantom{\dfrac{dM_i}{dt} =  \alpha_{i-1} M_{(i-1)}} \text{Model B} \\
& \dfrac{dM_i}{dt} = \alpha_{i-1} M_{(i-1)} - \alpha_i M_{i} + \mu N - \delta M_i \\
& \dfrac{dS_{1i}}{dt} = \alpha_{i-1} S_{1(i-1)} - \alpha_i S_{1i} + \delta M_i - \lambda_i S_{1i} \\
& \dfrac{dI_{1i}}{dt} = \alpha_{i-1} I_{1(i-1)} - \alpha_i I_{1i} + \lambda_i S_{1i} -  \gamma_1 I_{1i} \\
& \dfrac{dR_{1i}}{dt} =\alpha_{i-1} R_{1(i-1)} - \alpha_i R_{1i} +  \gamma_1 I_{1i} - \tau R_{1i} \\
& \dfrac{dS_{2i}}{dt} = \alpha_{i-1} S_{2(i-1)} - \alpha_i S_{2i} + \tau R_{1i} - \lambda_{2i} S_{2i} \\
& \dfrac{dI_{2i}}{dt} = \alpha_{i-1} I_{2(i-1)} - \alpha_i I_{2i} + \lambda_{2i} S_{2i} -  \gamma_{2i} I_{2i} \\
& \dfrac{dR_{2i}}{dt} = \alpha_{i-1} R_{2(i-1)} - \alpha_i R_{2i} + \gamma_2 I_{2i} - \tau R_{2i} \\
& \dfrac{dS_{3i}}{dt} = \alpha_{i-1} S_{3(i-1)} - \alpha_i S_{3i} + \tau R_{2i} + \tau R_{3i} - \lambda_{3i} S_{3i} \\
& \dfrac{dI_{3i}}{dt} = \alpha_{i-1} I_{3(i-1)} - \alpha_i I_{3i} + \lambda_{3i} S_{3i} -  \gamma_2 I_{3i} \\
& \dfrac{dR_{3i}}{dt} = \alpha_{i-1} R_{3(i-1)} - \alpha_i R_{3i} + \gamma_2 I_{3i} - \tau R_{3i}
\end{align*}

Here in addition to an initial period of maternal immunity, individuals can be in the susceptible, infected, or recovered classes for their first $(S_1, I_1, R_1)$, second $(S_2, I_2, R_2)$, or third and subsequent $(S_3, I_3, R_3)$ infections. The force of infection for age class $i$ is given by
$\lambda_i =  \displaystyle\sum_{j=1}^6 \beta_j(t) C_{ij} \dfrac{ (I_{1j} + 0.5 I_{2j} + 0.2 I_{3j}) }{ N_j }$,
assuming that relative infectiousness decreases for subsequent infections. We assume the relative risk of infection decreases for subsequent infections, setting $\lambda_{2i} = 0.62 \lambda_i$ and $\lambda_{3i} = 0.37 \lambda_i$ as in \cite{pitzer2012direct}. Only 13\% of first infections and 3\% of second infections are assumed to develop severe RVGE, based on data from a Mexico cohort study \citep{velazquez1996rotavirus}. So the expected number of reported cases for age class $i$ is given by $\rho \lambda_i (0.13 S_{1i} + 0.03 S_{2i})$. Following \cite{velazquez1996rotavirus}, we assume that the mean infectious period for the first infection is one week ($\gamma_1 = 1$) and for subsequent infections is half a week ($\gamma_2 = 2$).

Model C \citep{de2010dynamic} is an SEIRS model, similar to Model B but allowing for an additional exposed or incubation period. Individuals in the exposed class are infected but not yet infectious. The dynamic equations are given by:

\begin{align*}
& \phantom{\dfrac{dM_i}{dt} =  \alpha_{i-1} M_{(i-1)}} \text{Model C} \\
& \dfrac{dM_i}{dt} = \alpha_{i-1} M_{(i-1)} - \alpha_i M_{i} + \mu N - \delta M_i \\
& \dfrac{dS_{1i}}{dt} = \alpha_{i-1} S_{1(i-1)} - \alpha_i S_{1i} + \delta M_i - \lambda_i S_{1i} \\
& \dfrac{dE_{1i}}{dt} = \alpha_{i-1} E_{1(i-1)} - \alpha_i E_{1i} + \lambda_i S_{1i} -  \xi E_{1i} \\
& \dfrac{dI_{1i}}{dt} = \alpha_{i-1} I_{1(i-1)} - \alpha_i I_{1i} + \xi E_{1i} -  \gamma_1 I_{1i} \\
& \dfrac{dR_{1i}}{dt} =\alpha_{i-1} R_{1(i-1)} - \alpha_i R_{1i} +  \gamma_1 I_{1i} - \tau R_{1i} \\
& \dfrac{dS_{2i}}{dt} = \alpha_{i-1} S_{2(i-1)} - \alpha_i S_{2i} + \tau R_{1i}- \lambda_i S_{2i} \\
& \dfrac{dE_{2i}}{dt} = \alpha_{i-1} E_{2(i-1)} - \alpha_i E_{2i} + \lambda_i S_{2i} -  \xi E_{2i} \\
& \dfrac{dI_{2i}}{dt} = \alpha_{i-1} I_{2(i-1)} - \alpha_i I_{2i} + \xi E_{2i} -  \gamma_2 I_{2i} \\
& \dfrac{dR_{2i}}{dt} =\alpha_{i-1} R_{2(i-1)} - \alpha_i R_{2i} +  \gamma_2 I_{2i} - \tau R_{2i} \\
& \dfrac{dS_{3i}}{dt} = \alpha_{i-1} S_{3(i-1)} - \alpha_i S_{3i} + \tau R_{2i} - \lambda_i S_{3i} \\
& \dfrac{dE_{3i}}{dt} = \alpha_{i-1} E_{3(i-1)} - \alpha_i E_{3i} + \lambda_i S_{3i} -  \xi E_{3i} \\
& \dfrac{dI_{3i}}{dt} = \alpha_{i-1} I_{3(i-1)} - \alpha_i I_{3i} + \xi E_{3i} -  \gamma_2 I_{3i} \\
& \dfrac{dR_{3i}}{dt} =\alpha_{i-1} R_{3(i-1)} - \alpha_i R_{3i} +  \gamma_2 I_{3i} - \tau R_{3i}
\end{align*}

The modeling assumptions are the same as Model B but for the addition of an exposed class for the first, second, or subsequent infections $(E_1, E_2, E_3)$. We assume a mean exposed period of 1 day ($\xi = 7$).

Model D \citep{van2010mathematical} is an SIS model which also allows for successive infections with different levels of infectiousness, but assumes there is no period of temporary immunity following infection. After four infections individuals are assumed to be fully immune to infection. The dynamics are described as follows.

\begin{align*}
& \phantom{\dfrac{dM_i}{dt} =  \alpha_{i-1} M_{(i-1)}} \text{Model D} \\
& \dfrac{dM_i}{dt} = \alpha_{i-1} M_{(i-1)} - \alpha_i M_{i} + \mu N - \delta M_i \\
& \dfrac{dS_{1i}}{dt} = \alpha_{i-1} S_{1(i-1)} - \alpha_i S_{1i} + \delta M_i - \lambda_i S_{1i} \\
& \dfrac{dI_{1i}}{dt} = \alpha_{i-1} I_{1(i-1)} - \alpha_i I_{1i} + \lambda_i S_{1i} -  \gamma_1 I_{1i} \\
& \dfrac{dS_{2i}}{dt} = \alpha_{i-1} S_{2(i-1)} - \alpha_i S_{2i} + \gamma_1 I_{1i} - \lambda_{2i} S_{2i} \\
& \dfrac{dI_{2i}}{dt} = \alpha_{i-1} I_{2(i-1)} - \alpha_i I_{2i} + \lambda_{2i} S_{2i} -  \gamma_{2i} I_{2i} \\
& \dfrac{dS_{3i}}{dt} = \alpha_{i-1} S_{3(i-1)} - \alpha_i S_{3i} + \gamma_2 I_{2i} - \lambda_{3i} S_{3i} \\
& \dfrac{dI_{3i}}{dt} = \alpha_{i-1} I_{3(i-1)} - \alpha_i I_{3i} + \lambda_{3i} S_{3i} -  \gamma_2 I_{3i} \\
& \dfrac{dS_{4i}}{dt} = \alpha_{i-1} S_{4(i-1)} - \alpha_i S_{4i} + \gamma_2 I_{3i} - \lambda_{4i} S_{4i} \\
& \dfrac{dI_{4i}}{dt} = \alpha_{i-1} I_{4(i-1)} - \alpha_i I_{4i} + \lambda_{4i} S_{4i} -  \gamma_2 I_{4i}
\end{align*}

The force of infection is $\lambda_i =  \displaystyle\sum_{j=1}^6 \dfrac{ \beta_j(t) C_{ij} (I_1 + 0.5 I_2 + 0.2 I_3 + 0.2 I_4) }{ N }$, assuming that relative infectiousness decreases for subsequent infections. We also assume the relative risk of infection decreases for subsequent infections, setting $\lambda_{2i} = 0.62 \lambda_i$ and $\lambda_{3i} = \lambda_{4i} = 0.37 \lambda_i$. Again, we assume only 13\% of first infections and 3\% of second infections are assumed to develop severe RVGE. So the expected number of reported cases in age group $i$ is given by $\rho \lambda (0.13 S_{1i} + 0.03 S_{2i})$.

Finally, Model E \citep{atchison2010modelling} is an SIR-SIS hybrid wherein following infection, individuals have a chance to either return to the susceptible class or gain full immunity. The equations for the dynamics are as follows. 

\begin{align*}
& \phantom{\dfrac{dM_i}{dt} =  \alpha_{i-1} M_{(i-1)}} \text{Model E} \\
& \dfrac{dM_i}{dt} = \alpha_{i-1} M_{(i-1)} - \alpha_i M_{i} + \mu N - \delta M_i \\
& \dfrac{dS_{1i}}{dt} = \alpha_{i-1} S_{1(i-1)} - \alpha_i S_{1i} + \delta M_i - \lambda_i S_{1i} \\
& \dfrac{dI_{1i}}{dt} = \alpha_{i-1} I_{1(i-1)} - \alpha_i I_{1i} + \lambda_i S_{1i} -  \gamma_1 I_{1i} \\
& \dfrac{dS_{2i}}{dt} = \alpha_{i-1} S_{2(i-1)} - \alpha_i S_{2i} + \kappa_1 \gamma_1 I_{1i} - \lambda_{2i} S_{2i} \\
& \dfrac{dI_{2i}}{dt} = \alpha_{i-1} I_{2(i-1)} - \alpha_i I_{2i} + \lambda_{2i} S_{2i} -  \gamma_{2i} I_{2i} \\
& \dfrac{dS_{3i}}{dt} = \alpha_{i-1} S_{3(i-1)} - \alpha_i S_{3i} + \kappa_2 \gamma_2 I_{2i} - \lambda_{3i} S_{3i} \\
& \dfrac{dI_{3i}}{dt} = \alpha_{i-1} I_{3(i-1)} - \alpha_i I_{3i} + \lambda_{3i} S_{3i} -  \gamma_2 I_{3i} \\
& \dfrac{dS_{4i}}{dt} = \alpha_{i-1} S_{4(i-1)} - \alpha_i S_{4i} + \kappa_3 \gamma_2 I_{3i} - \lambda_{4i} S_{4i} \\
& \dfrac{dI_{4i}}{dt} = \alpha_{i-1} I_{4(i-1)} - \alpha_i I_{4i} + \lambda_{4i} S_{4i} -  \gamma_2 I_{4i}
\end{align*}

The chance of returning to the susceptible class varies by number of previous infections. Following \cite{atchison2010modelling} we fix $\kappa_1 = 0.62$, $\kappa_2 = 0.65$, $\kappa_3 = 0.85$. The remaining modeling assumptions are the same as for models B-D.

\subsection{Computational Details}

~~~~Denote the observed data by $C = \{ C_{i,t}; t \in (1,...,t_{obs}), i \in (1,...,6)\}$ where $C_{i,t}$ is the number of reported cases in age group $i$ during week $t$. Cases were observed over $t_{obs}=118$ weeks. Denote the number of cases in age group $i$ during week $t$ predicted by our models by $\xi_{i}(t)$. For Model A,
\[
\xi_{i}(t) = 0.24 \rho \lambda_{i,t} S_{i,t}
\]
While for models B-E,
\[
\xi_{i}(t) = \rho \lambda_{i,t} (0.13 S_{1i,t} + 0.03 S_{2i,t})
\]

For each model, the periodic solution to the system of ODEs specified above determines the number of reported cases in age group $i$ during a given week. Model dynamics are integrated forward using the \texttt{deSolve} \citep{deSolve} package in \texttt{R} until a periodic solution is reached. Solutions have a period of one year; that is, starting from arbitrary initial conditions, we run the dynamics forward until our expected number of cases is identical from one 52 week period to the next, to within a small tolerance; i.e.
\[
\sum_{i=1}^6 \sum_{t=t^*}^{t^*+52} | \xi_i(t) - \xi_i(t-52) | < \epsilon = 0.01
\]
In practice, numerical integration for 20 years was enough to ensure the periodic solution was reached. After reaching a periodic solution, the models are integrated forward an additional 118 weeks to get the expected number of reported cases $\left( \Xi_{i}(t); t \in (1,...,t_{obs}), i \in (1,...,6) \right)$.

Define random variables $N_{i,t} \sim NB(\Xi_i(t), r)$. The likelihood is
\[
\mathcal{L} (C| \Theta) = \prod_{i=1}^6\prod_{t=1}^{t_{obs}} f_{N_{i,t}}( C_{i,t} )
\]
The number of observed reported cases is modeled as a Negative Binomial with mean equal to the expected number of cases and dispersion parameter $r$.

Inference for our model parameters is done via Markov chain Monte Carlo (MCMC) for models A-E. At each step of the Markov chain, new parameters $\Theta'$ are proposed and the model dynamics are integrated forward until the periodic solution $\Xi_{i}(t;\Theta')$ is reached in order to calculate $\mathcal{L} (C| \Theta')$. The parameters estimated by MCMC are $\Theta = \left( b,\phi,r,\rho,\beta_{0i}; i \in (1,...,6) \right)$, including seasonal amplitude $b$, seasonal phase $\phi$, the dispersion $r$ of the Negative Binomial observation process, the reporting rate $\rho$, and the baseline transmission rate for age class $\beta_{0i}$.

MCMC samples are obtained from the posterior distribution
\[
\pi( b,\phi,r,\rho,\beta_{01},...,\beta_{06} | C) \propto \mathcal{L} (C | b,\phi,r,\rho,\beta_{01},...,\beta_{06}) p(b) p(\phi) p(r) p(\rho) \prod_{i=1}^6 p(\beta_{0i})
\]

where we take priors $p(\beta_{0i}) = N(20,5)$, $p(b) = \text{Unif}(0,1)$, $p(\phi) = \text{Unif}(2,2\pi + 2)$, $p(r) = \text{Gamma}(0.001,0.001)$, and $p(\rho) = N(0.117,0.06)$. The prior of our reporting rate $\rho$ is centered at 11.7\%, determined from the estimated reporting rate from the cluster survey (42.9\%) and the estimated proportion of the population under 5 in the four districts that is covered by hospital surveillance (27.3\%, from 2009 census data). In practice, we find that our estimates are robust to the choice of standard deviation of $p(\rho)$.
 
Table \ref{tab:postmeans} provides parameter estimates from five different models. Estimates of the strength of transmission are similar for models B-D, higher for Model E and significantly lower for Model A. The same holds true for the reporting rate (Model A's estimate of the reporting rate is dramatically lower, and does  not agree with estimates from the cluster survey, evidence that it is performing poorly). Notably, the estimated phase of the transmission $\phi$ is similar across all models (Table \ref{tab:postmeans}). For reference, an estimated $\phi$ of $7.4$ corresponds to a peak transmission in early March. This is quite close to the period of peak night time brightness in Maradi as measured by satellite imagery \citep{bharti2011explaining}.  
 
\begin{table}
\centering
\caption{Posterior means and 95\% HPD intervals}
\begin{tabular}{c|ccccc}
\hline
Model & $b$ & $\phi$ & $r$ & $\rho$ \\
\hline
A & 0.50 (0.48,0.51) & 7.4 (7.3,7.5) & 1.5 (1.4,1.5) & 0.039 (0.035,0.044) \\
B & 0.41 (0.37,0.45) & 7.4 (7.3,7.6) & 2.6 (2.3,2.8) & 0.096 (0.089,0.104) \\
C & 0.42 (0.36,0.46) & 7.4 (7.3,7.5) & 2.5 (2.0,2.8) & 0.097 (0.089,0.104) \\
D & 0.30 (0.26,0.34) & 7.2 (7.0,7.4) & 2.7 (2.5,2.8) & 0.098 (0.090,0.107) \\
E & 0.32 (0.27,0.36) & 7.1 (7.0,7.3) & 2.6 (2.5,2.7) & 0.109 (0.099,0.117) \\
\hline
\end{tabular}
\label{tab:postmeans}
\end{table}

\subsection{Dynamics Accounting for Vaccination}

~~~~ Based on the results of \citep{lopman2012understanding} we assume that $63\%$ of vaccinated individuals are successfully seroconvert after a single dose. Our models with vaccination allow for the red transitions in Figure \ref{modelstructure}. For example, Model B allows for transitions directly from $M_{i=1}$ and $S_{1,i=1}$ to $R_{1,i=2}$ on the first dose, and from $R_{1,i=2}$ and $S_{2,i=2}$ to $R_{2,i=3}$ on the second dose. The dynamics equations will be modified by the following terms:

\begin{align*}
& \dfrac{dM_{i=2}}{dt} = (1 - s c) \alpha_1 M_{i=1} + ...\\
& \dfrac{dS_{1,i=2}}{dt} = (1 - s c) \alpha_1 S_{1,i=1} + ...\\
& \dfrac{dR_{1,i=2}}{dt} = (s c) \alpha_1 M_{i=1} + (s  c) \alpha_1 S_{1,i=1} + ...\\
& \dfrac{dR_{1,i=3}}{dt} = (1 - s  c) \alpha_2 R_{1,i=2} + ...\\
& \dfrac{dS_{2,i=3}}{dt} = (1 - s  c) \alpha_2 S_{2,i=2} + ...\\
& \dfrac{dR_{2,i=3}}{dt} = (s  c) \alpha_2 R_{1,i=2} + (s  c) \alpha_2 S_{2,i=2} + ...
\end{align*}

Where $c$ is the coverage and $s=0.63$ is the rate of seroconversion \citep{lopman2012understanding} for low socio-economic settings.  This means that an individual who is vaccinated with a single dose has a lower risk of infection, comparable to the effect of recovering from a natural infection. Vaccination with a second dose further reduces risk of infection.

In Model A, the risk of infection does not decline with the previous number of infections. Therefore, an additional vaccinated state $V_i$ is added to the model for age group $i$. Two additional input parameters are required for the vaccine efficacy against severe and mild RVGE. We assume the vaccination happens at $2$ months, but the vaccine efficacy is equal to the efficacy predicted under models B-E for the two dose strategy, $\eta_s = .796$ and $\eta_m = .609$.

\begin{align*}
& \dfrac{dM_{i=2}}{dt} = (1 - c) \alpha_1 M_{i=1} + ...\\
& \dfrac{dS_{1,i=2}}{dt} = (1 - c) \alpha_1 S_{i=1} + ...\\
& \dfrac{dV_{i=2}}{dt} = (c) \alpha_1 M_{i=1} + (c) \alpha_1 S_{i=1} + ...\\
& \dfrac{dV_{i>2}}{dt} = \alpha_{i-1} V_{(i-1)} - \alpha_i V_{i} - (\tau + \lambda_{s,i} (1-\eta_s) + \lambda_{m,i} (1-\eta_m)) V_{i} \\
& \dfrac{dS_{i>2}}{dt} = \tau V_{i} + ...\\
& \dfrac{dI_{s,i>2}}{dt} = \lambda_{s,i} (1-\eta_s) V_i + ...\\
& \dfrac{dI_{m,i>2}}{dt} = \lambda_{m,i} (1-\eta_m) V_i + ...
\end{align*}

Given our vaccination strategy for models B-E, the vaccine efficacy for severe RVGE after two doses is 79.6\%, in line with efficacy studies of rotavirus vaccines. This is calculated by multiplying the proportion of individuals who are successfully immunized twice, once, or zero times by the expected reduction in RVGE incidence for each case.
\begin{equation}
VE = 1 - \left[ 0.37^2 \times 1 + 2 (0.37)(0.63) \times \left(0.62 \dfrac{0.03}{0.13} \right) + 0.63^2 \times \left(0.37 \dfrac{0}{0.13} \right) \right] = 79.6\%.
\label{ADDREF}
\end{equation}
We assume following \cite{velazquez1996rotavirus} that 47\% of first infections and 25\% of second infections and 32\% of third infections are assumed to develop any RVGE (mild RVGE is unreported). Therefore the vaccine efficacy for all RVGE is
\[
VE = 1 - \left[ 0.37^2 \times 1 + 2 (0.37)(0.63) \times \left(0.62 \dfrac{0.25}{0.47} \right) + 0.63^2 \times \left(0.37 \dfrac{0.32}{0.47} \right) \right] = 60.9\%.
\]

In practice, first model parameters $\Theta$ are estimated via MCMC for the models without vaccination. Using the fitted model, the dynamics are then integrated forward at the posterior mean of $\Theta$ until the periodic solution has been reached. Then, the dynamics are modified to allow for transitions between compartments by vaccination.

\subsubsection{Projections Based on Vaccine Efficacy from a Recent Study}

~~~~Recently \cite{isanaka2017efficacy} estimated that 3 doses of vaccine had 66.7\% efficacy against severe RVGE among children in Niger. Though we do not explicitly account for 3 doses of vaccine, we can calculate the effective seroconversion rate for our model above that would yield this observed efficacy after a complete sequence of doses. Thus, we set $\eta_s = .667$ and use \eqref{ADDREF} to calculate the effective seroconversion rate as 49\%. Then the vaccine efficacy for all RVGE is $\eta_m = .515$. We estimate the predicted impact of vaccination using different $\eta_s, \eta_m$ and $s$ values with the same dynamic equations. 

\begin{figure}
\centering
\includegraphics[width=.7\linewidth]{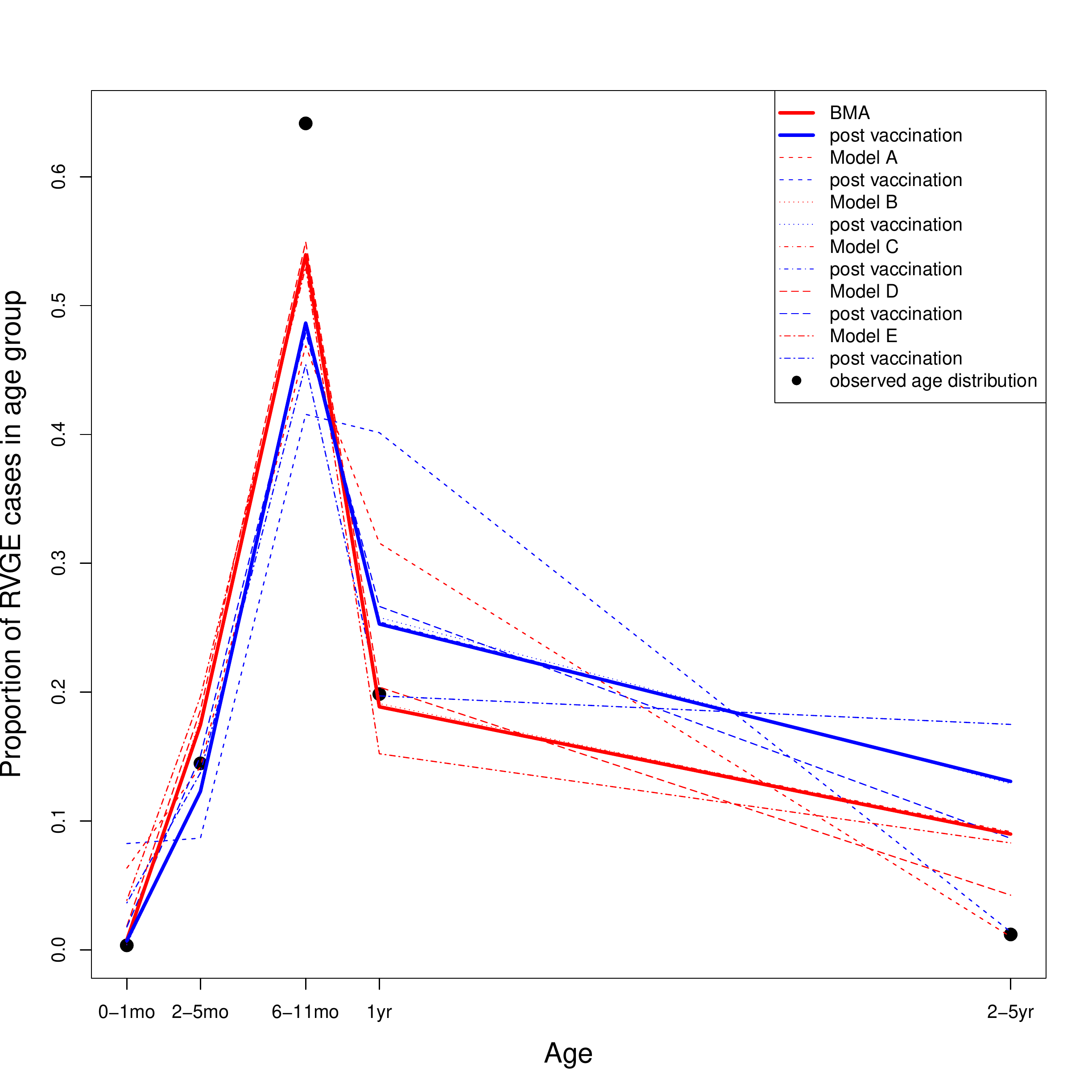}
\caption{Distribution of cases across age groups observed in the data (black dots), predicted by the models (red lines), and predicted after vaccination has been introduced at 70\% coverage (blue lines).}
\label{fig:NewBMAageDisttry2}
\end{figure}

Although we used the two dose strategy, by using different value of the efficacy, our study can account for uncertainty in the seroconversion rate. Figures 7-9 are matched to Figures 3-5 in the main paper. Because of the lower seroconversion rate, the projected results were qualitatively similar, quantitatively smaller. Vaccination causes a shift in the age distribution across models (Figure \ref{fig:NewBMAageDisttry2}), with a higher proportion of RVGE cases occurring for older children.

\begin{figure}
\includegraphics[width=\linewidth]{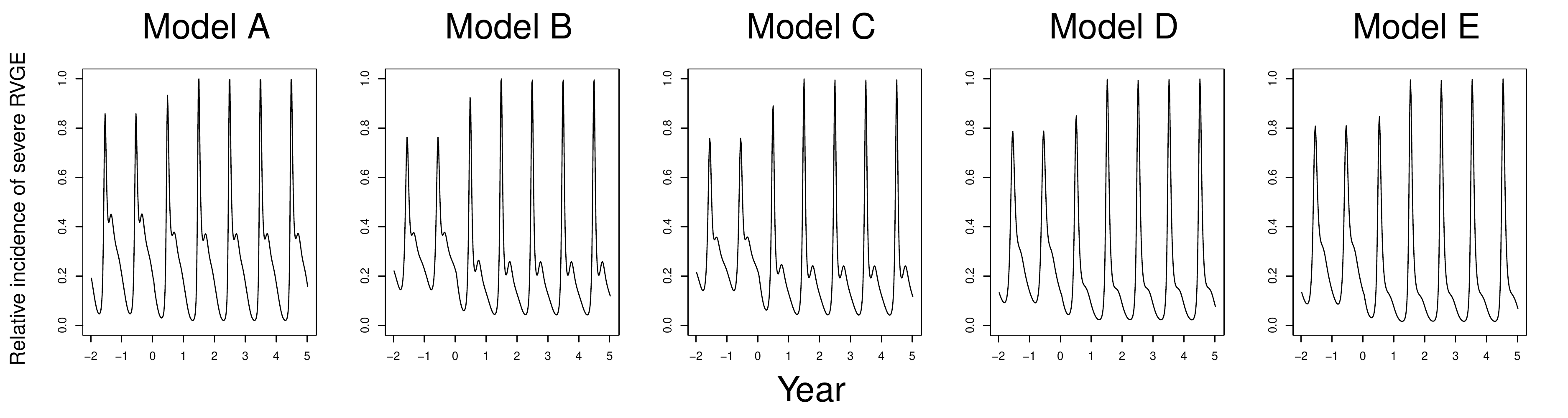}
\caption{Relative incidence of severe RVGE after vaccination has been introduced into the models assuming 70\% coverge, out to five years after vaccination has been introduced. The vaccination has been introduced at 0 year. }
\label{fig:NewvaccShortTerm}
\end{figure}

Over the short term, Models A-E predict an overall decline in total burden, but an increase in the magnitude of peak incidence (Figure \ref{fig:NewvaccShortTerm}).

\begin{figure}
\includegraphics[width=\linewidth]{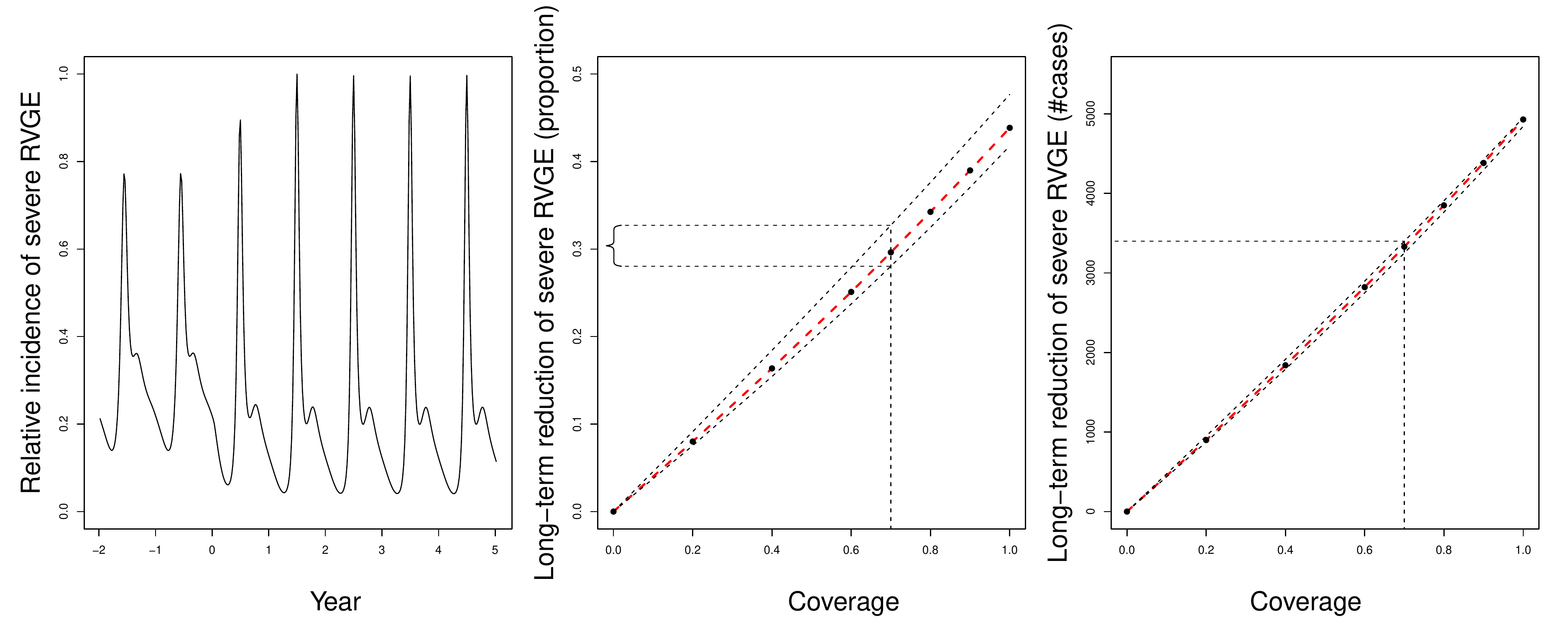}
\caption{Relative incidence of severe RVGE (Left), percent (Middle) and absolute (Right) long term reduction in cases by coverage for Bayesian model averaging from the five fitted models. Dashed lines denote 99\% confidence interval. The vaccination has been introduced at 0 year. Variation in reduction for a fixed (70\%) level of coverage is demonstrated.}
\label{fig:NewBMAvaccReduce}
\end{figure}

Figure \ref{fig:NewBMAvaccReduce} indicates that the short term trend of vaccination impacts based on BMA is similar to that of Model C. BMA predicts 29.6\% of long term reduction $\left(99\% \text{CI}: (28.0\%,32.7\%)\right)$.

\section{Bayesian Model Averaging}

~~~~For $k=1,...,5$, consider $M_{k}$, the $k$th model, with prior $p(\Theta_{k}|M_{k})$ and  likelihood function $\mathcal{L}(C|\Theta_{k},M_{k})$. Note that we take the uniform model prior for $p(M_{l})$ and model evidence $P(C|M_{k})$ is approximated via Bayesian information criterion (BIC) as in \cite{raftery1996approximate}. Then the posterior model probability (PMP) for $M_{k}$ given the observed data $C$ is
\[
p(M_{k}|C)=\frac{p(C|M_{k})p(M_{k})}{\sum_{l=1}^{5}p(C|M_{l})p(M_{l})},
\]
where
\[
p(C|M_{k})=\int \mathcal{L}(C|\Theta_{k},M_{k})p(\Theta_{k}|M_{k})d\Theta_{k}
\]
is the model evidence for $M_{k}$ which measures how well each model is supported by the observed data. Then the BMA estimate of the burden is 
\[
E[\xi(t)|C] = \sum_{l=1}^{5} E[\xi_{l}(t)|C,M_{l}]p(M_{l}|C).
\]

A summary of our implementation of BMA is as follows: (1) We construct a separate MCMC algorithm for each of the models A-E. (2) For each model, the burden estimate $\xi_{k}(t)$ is evaluated for the MCMC samples of the posterior distribution of that model. (3) The expected burden for model $k$, $E[\xi_{k}(t)|C,M_{k}]$, is estimated through the sample mean of the  $\xi_{k}(t)$s obtained from Step (2). (4) We take the weighted average of the burden across all models, with the weights equal to the posterior model probabilities, $p(M_{k}|C)$, obtained above.

\bibliographystyle{plain}
\bibliography{references}

\end{document}